\documentclass[lettersize,journal]{IEEEtran}
\ifCLASSINFOpdf
   \usepackage[pdftex]{graphicx}
\else
   \usepackage[dvips]{graphicx}
\fi

\usepackage[T1]{fontenc}
\usepackage[cmex10]{amsmath}
\usepackage{amsfonts}

\ifCLASSOPTIONcompsoc
  \usepackage[caption=false,font=normalsize,labelfont=sf,textfont=sf]{subfig}
\else
  \usepackage[caption=false,font=footnotesize]{subfig}
\usepackage{subcaption}
\usepackage{fancyhdr}
\usepackage{multicol}
\usepackage{url}
\usepackage[colorlinks=true, allcolors=blue]{hyperref}
\usepackage{wrapfig}
\usepackage{tikz}
\usepackage{pgfplots}
\pgfplotsset{compat=1.12}
\usepgfplotslibrary{fillbetween}
\usetikzlibrary{external,
                arrows.meta,
                fit,
                positioning,
                calc,
                patterns,
                decorations.pathmorphing,
                shapes.geometric}
\usepackage{amsthm}
\usepackage{eurosym}                

\theoremstyle{definition}

\usepackage{nomencl}
\makenomenclature
\usepackage{etoolbox}
\renewcommand\nomgroup[1]{%
  \item[\noindent\bfseries
  \ifstrequal{#1}{D}{Primal variables}{%
  \ifstrequal{#1}{E}{Dual variables}{%
  \ifstrequal{#1}{C}{Parameters}{%
  \ifstrequal{#1}{B}{Sets}{%
  \ifstrequal{#1}{A}{Functions}{}
  }}}}%
]}
\makeatletter
\let\old@ps@headings\ps@headings
\let\old@ps@IEEEtitlepagestyle\ps@IEEEtitlepagestyle
\def\psccfooter#1{%
    \def\ps@headings{%
        \old@ps@headings%
        \def\@oddfoot{\strut\hfill#1\hfill\strut}%
        \def\@evenfoot{\strut\hfill#1\hfill\strut}%
    }%
    \def\ps@IEEEtitlepagestyle{%
        \old@ps@IEEEtitlepagestyle%
        \def\@oddfoot{\strut\hfill#1\hfill\strut}%
        \def\@evenfoot{\strut\hfill#1\hfill\strut}%
    }%
    \ps@headings%
}
\makeatother

\begin{document}
\onecolumn
This work has been submitted to the IEEE for possible publication. Copyright may be transferred without notice, after which
this version may no longer be accessible.
\pagenumbering{gobble}
\newpage
\pagenumbering{arabic}
\twocolumn
\title{Refinement of Reliability Grid Codes in \\ the Provision of Ancillary Services}

% To specify the authors when (number of affiliations > 2)
\author{\IEEEauthorblockN{Torine R. Herstad,
Jalal Kazempour, \IEEEmembership{Senior Member,  IEEE}, \\
Lesia Mitridati, and
Steven A. Gabriel, \IEEEmembership{Senior Member,  IEEE} \vspace{-2em}}
%\thanks{Manuscript was received on 2026-06-01.}
\thanks{Torine R. Herstad, Jalal Kazempour, and Lesia Mitridati are with the Technical University of Denmark, Kgs. Lyngby, Denmark (e-mails: \{torhe, jalal, lemitri\}@dtu.dk).}
\thanks{Steven A. Gabriel is with the University of Maryland, MD, USA, the Norwegian University of Science and Technology, Trondheim, Norway, and Aalto University, Espoo, Finland (e-mail: sgabriel@umd.edu).
}}

% make the title area
\maketitle
\pagestyle{fancy}
\fancyhf{}
\fancyhead[R]{\thepage}

\begin{abstract}
Stochastic resources such as wind farms, electric vehicle aggregators, and demand-side assets are increasingly participating as reserve providers in ancillary service markets. To manage delivery uncertainty, system operators impose minimum reliability thresholds on such providers. Energinet, the Danish transmission system operator (TSO), has pioneered this approach through the P90 requirement, requiring stochastic providers to make accepted reserve capacity bids available with at least 90\% probability. Yet this threshold is set by regulatory convention, not optimization: no existing framework treats it as a design variable or characterizes the cost-reliability trade-off it governs. This paper closes that gap. We develop a bilevel optimization framework in which the TSO in the upper level sets the reliability threshold endogenously while providers in the lower levels respond through reliability-constrained bidding, with chance constraints reformulated analytically using a Weibull tail distribution. Applied to the Nordic frequency containment reserve for disturbances (FCR-D) market, the cost-optimal threshold lies below P90 in the studied cases, with cost reductions by up to 14.5\% relative to the fixed standard. Dynamic hourly thresholds yield a further reduction of up to 2.4\%, suggesting efficiency gains may increase in larger and more diverse reserve markets.
\end{abstract}

\vspace{-1mm}

\vspace{-2mm}
\printnomenclature
\section{Introduction}

Maintaining grid stability requires transmission system operators (TSOs) to procure sufficient ancillary services, including reserves for balancing purposes\footnote{Hereafter, we use the terms ancillary services and reserves interchangeably, although ancillary services generally encompass a broader range of services beyond reserves.}. This reserve procurement task is becoming more complex as wind, solar, and demand-side flexible assets such as electric vehicles (EVs) contribute to the reserve provision. Unlike dispatchable generation, these assets are inherently stochastic, challenging the procurement assumptions that reserve market design has historically relied upon.

Early reserve procurement frameworks were designed around a fully reliable supply base of dispatchable thermal and hydro generators. In this setting, operational uncertainty arose from demand forecast errors, unplanned generation outages, and transmission contingencies. Reserve dimensioning was therefore grounded in deterministic security criteria, most notably the N-1 rule, which required sufficient upward reserve to cover the loss of the single largest generator at all times. Seminal work from this period established the theoretical and computational foundations for optimal reserve sizing under these conditions, consistently arriving at the conclusion that reserves should be sourced exclusively from controllable generators and dimensioned to guarantee full coverage \cite{Wood1996PowerGO}.

The large-scale integration of wind power from the early 2000s onward fundamentally altered this picture. Wind generation forecast errors introduced an additional stochastic component into net demand, and empirical studies documented a corresponding increase in required reserve volumes \cite{6039157, FRUNT20101528,7999262}. Despite this shift, reserve provision remained the exclusive domain of dispatchable generators. Wind and solar were treated as exogenous disturbances to be balanced rather than as potential contributors to system flexibility, and the assumption that reserves must be sourced from fully controllable units remained intact even as the uncertainty those reserves were designed to cover grew substantially \cite{4682642}.

This convention is now being challenged. As the share of stochastic resources grows, excluding them from reserve provision becomes increasingly costly. Consequently, system operators are beginning to allow stochastic renewable resources to participate directly as reserve providers \cite{https://doi.org/10.1049/iet-gtd.2014.0614}. Demand-side assets are also increasingly being integrated into reserve provision frameworks \cite{hrvoje,mads}. A small but growing literature has examined this possibility, typically by applying the same participation rules to stochastic reserve providers as to dispatchable ones \cite{DIAZGONZALEZ2014551}. However, this approach does not address how participation rules themselves should be designed for inherently stochastic assets. Imposing conventional grid codes on them either excludes cheaper providers who cannot meet a full reliability standard, or exposes the system to reserve shortfall risk if standards are relaxed without formal justification. This is precisely the gap that the \textit{P90 requirement} of the Danish TSO represents a first attempt to fill.

Energinet, the Danish TSO, has pioneered a middle path through the P90 requirement, which permits stochastic reserve providers to participate while imposing a minimum delivery probability of 90\% \cite{Energinet}. In practice, Energinet pre-qualifies reserve providers to bid into the reserve capacity market and maintains their qualification through an \textit{ex-post} check: over a rolling three-month evaluation period, providers must demonstrate that their historical reserve capacity bids were available at least 90\% of the time, regardless of whether the reserve was actually activated, based on their realized stochastic profiles including historical production for wind and solar and realized flexibility for demand-side aggregators. This has enabled stochastic renewable resources and demand-side flexibility aggregators to enter the Nordic day-ahead reserve capacity markets. However, the 90\% threshold itself remains static and \textit{ad hoc}, set by regulatory convention rather than derived from any explicit optimization. No existing framework treats the reliability threshold as a design variable or formally characterizes the Pareto-optimal trade-off between procurement cost and delivery reliability that the threshold governs.

This paper addresses that gap directly. We develop a bilevel optimization framework in which the TSO sets the reliability threshold endogenously by balancing reserve procurement costs against the expected cost of under-delivery, the so-called reserve shortfall, modeling the strategic response of a heterogeneous portfolio of reserve providers including renewables, demand-side aggregators, and dispatchable generators. Uncertainty is handled through chance constraints, analytically reformulated using a Weibull tail distribution for computational tractability. The specific contributions are:

\begin{itemize}
\item We formulate reliability grid code design as a bilevel optimization problem, with the TSO's threshold as the upper-level decision variable and reserve provider bidding and reserve market clearing as the lower-level response.

\item We demonstrate that exogenously setting static reliability thresholds, such as the current P90 standard, is generally not cost-optimal, and that endogenously deciding on the reliability can reduce total reserve provision costs up to 14.5\%, while allowing for time-varying dynamic thresholds can further reduce costs up to 2.4\%.

\end{itemize}

The remainder of the paper is organized as follows. Section II describes the Nordic reserve capacity market context and the P90 requirement. Section III presents the mathematical model. Section IV describes the case study and results. Section V concludes and discusses directions for future work.

\section{Preliminaries and Our Focus}
\label{prelim}

The Nordic reserve markets comprise multiple auctions across different products and timeframes. All reserve products are traded in day-ahead capacity markets, while some are additionally settled in energy activation markets closer to real time. Products differ in activation characteristics including response speed and energy volume, ranging from near-instantaneous frequency containment to slower restoration reserves. In the day-ahead capacity market, the TSO does not yet know the supply mix, the level of stochasticity at delivery, or the reliability with which procured reserves can be provided.

In Denmark, Energinet procures the necessary reserve volumes to maintain grid stability in coordination with the Swedish TSO. Stochastic reserve providers, such as wind power aggregators and EV aggregators, must comply with the P90 requirement, which requires that offered capacity be available with at least 90\% probability. Compliance is assessed ex-post over a rolling three-month evaluation period: if a provider's unavailability exceeds 10\%, it risks losing market participation eligibility. This availability check applies regardless of whether activation was actually requested. In addition, if a provider is activated and fails to deliver, it bears the cost of the alternative resource procured to cover the shortfall. These two mechanisms together, eligibility risk from persistent unavailability and cost exposure upon activation failure, define the consequence structure that stochastic providers face under the P90 requirement.

Among the diverse reserve products in the Nordic markets, less energy-intensive and less frequently activated services are particularly attractive to stochastic providers such as wind farms and EV aggregators. Frequency Containment Reserve for Disturbances (FCR-D), available across the Nordic synchronous area including eastern Denmark, Sweden, Norway, and Finland \cite{FrequencyControlOverviewENTSOE}, is a prime example. Since FCR-D is not energy-intensive and grid frequency deviates beyond its activation thresholds of 49.9 Hz and 50.1 Hz only between zero and a few hours per day \cite{GandE}, it does not materially compromise battery charging schedules, making it an attractive opportunity for EV fleets and other battery-driven units \cite{GandE}. Wind farms can similarly benefit, since blade pitching enables near-instantaneous output adjustments \cite{7525114}. FCR-D is an asymmetric product with separate markets for up- and down-regulation, and providers are compensated primarily through capacity reservations settled at marginal pricing. Activated energy is settled through the standard imbalance mechanism, and given the intermittent nature of FCR-D activations, the associated imbalance costs are small relative to capacity revenues. This makes FCR-D a natural and practically relevant setting for studying the P90 requirement, and this paper therefore treats it as a capacity market for modeling purposes.

\begin{figure}[t]
  \begin{center}
\begin{tikzpicture}[
        node distance = 4mm and 8mm,
            N/.style = {draw, rounded corners, fill=white, minimum size=1.2cm, align=center},
            b_arr/.style = {->, rounded corners=1mm},
            punkt/.style={draw, rectangle, rounded corners, fill=white, thick, inner sep=2.5pt, align=center},
            every edge/.append style = {draw, semithick, -Stealth}
                                ]
        % UPPER LEVEL
        \node[N, ultra thick, minimum width=2.8in] (TSO) {Transmission system operator\\ \small{\textit{minimize overall reserve-related cost}}};
        
        % LOWER LEVELS
        % EV
        \node[punkt, minimum width=3.5cm, below=1.2cm of TSO.west, anchor=north west] (EV) {\begin{tikzpicture}
                \node[] (EV_text) {EV aggregator};
                \node[below=0pt of EV_text] (EV_obj) {\small{\textit{maximize bid size}}};
            \end{tikzpicture}};
        % WIND
        \node[punkt, minimum width=3.5cm, below=0.3cm of EV.south, anchor=north] (wind) {\begin{tikzpicture}
                \node[] (wind_text) {Wind farm};
                \node[below=0pt of wind_text] (wind_obj) {\small{\textit{maximize bid size}}};
            \end{tikzpicture}};
        % CONVENTIONAL GENERATOR
        \node[punkt, minimum width=3.5cm, below=0.3cm of wind.south, anchor=north] (conv) {\begin{tikzpicture}
                \node[] (text) {Dispatchable generator};
                \node[below=0pt of text] (obj) {\small{\textit{maximize bid size}}};
            \end{tikzpicture}};
        % MARKET
        \node[punkt, minimum width=2.5cm, below=1.2cm of TSO.east, anchor=north east] (market) {\begin{tikzpicture}
                \node[] (market_text) {Market clearing};
                \node[below=0pt of market_text] (market_obj) {\small{\textit{minimize cost}}};
            \end{tikzpicture}};
        % ARROWS
        \draw[<-] (EV.north) -- node[pos=0.5, right] {$1-\varepsilon_t$} ++(0,0.6cm);
        \draw[b_arr] (TSO.west) -- ++(-0.3,0)
    -- (wind.west -| {$(TSO.west)+(-0.3,0)$})
    node[midway, above, rotate=90] {$1-\varepsilon_t$}
    |- (wind.west);
        \draw[->] ([shift={(0.3,0)}]market.north) -- node[pos=0.5, right] {$b_{t,i}$} ++(0,0.6cm);
        \draw[<-] ([shift={(-0.3,0)}]market.north) -- node[pos=0.5, left] {$d_t$} ++(0,0.6cm);
        \draw[->] (EV.east) -- node[pos=0.5, above] {$\Hat{b}_{t,i}$} (market.west);
        \draw[b_arr] (wind.east) -| node[pos=0.1, above] {$\Hat{b}_{t,i}$} ([shift={(-0.3,0)}]market.south);
        \draw[b_arr] (conv.east) -| node[pos=0.1, above] {$\Hat{b}_{t,i}$} ([shift={(0.3,0)}]market.south);
        \end{tikzpicture}
\end{center}
\caption{\small{Information exchange in the bilevel optimization formulation. 
In the upper level, the TSO sets the reliability threshold $1-\varepsilon_t$, 
optimized hourly in the dynamic case and reduced to a single value 
$1-\varepsilon$ in the static case, to minimize total reserve-related costs, 
including reserve procurement and expected shortfall penalties. In the lower 
level, a representative EV aggregator and a representative wind farm
receive the threshold and solve convex chance-constrained problems to maximize 
their reserve bids $\hat{b}_{t,i}$ based on probabilistic forecasts, while a 
representative dispatchable generator submits bids constrained only by 
available capacity. The reserve market-clearing problem then determines the 
cleared reserve capacities $b_{t,i}$ given reserve demand $d_t$ and all 
submitted bids, which are anticipated by the TSO when determining the optimal 
reliability threshold.}}
\vspace{-4mm}
\label{fig:infoex1}
\end{figure}
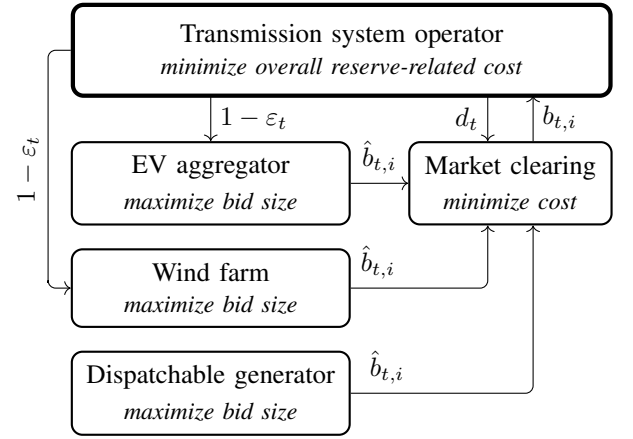

This paper focuses on the FCR-D up-regulation market as a tractable starting point for analyzing reliability grid code optimization. Beyond the capacity-dominated compensation structure discussed above, FCR-D is particularly well-suited because providers face no significant rebound effect following activation: an EV aggregator reducing its charging rate does not need to compensate with increased consumption afterward, and a wind farm adjusting output through blade pitching faces no comparable recovery obligation. This means the stochastic bid modeled here fully characterizes provider behavior without requiring a multi-period energy model. Extending the framework to energy-intensive services such as automatic Frequency Restoration Reserve (aFRR) and manual Frequency Restoration Reserve (mFRR), where rebound effects and multi-period energy constraints become relevant, is left as an important direction for future work.

\section{Mathematical model}\label{sec:mathematical model}
\nomenclature[B]{\(\mathcal{T}\)}{Hours, $t\in \mathcal{T}=\{1,\dots,\vert \mathcal{T}\vert=24\}$}
\nomenclature[B]{\(\mathcal{I}\)}{Reserve providers, $i\in I=\{1,\dots, \vert \mathcal{I}\vert\}$}
\nomenclature[B]{\(\Omega\)}{Scenarios, $\omega\in \Omega=\{1,\dots,\vert \Omega\vert\}$}
\nomenclature[Ba]{\(\Omega'\)}{Out-of-sample scenarios, $\omega'\in \Omega'=\{1,\dots,\vert \Omega'\vert\}$}

\begin{figure}[t]
    \centering
    \includegraphics[scale=1]{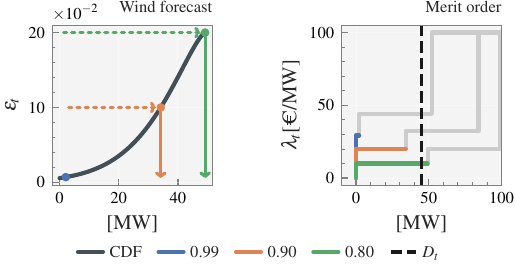}
    \caption{\small{Illustration of how the reliability threshold affects the wind farm's 
reserve bid and the resulting reserve merit order for a representative 
hour. The left plot shows the cumulative distribution function (CDF) of 
the wind power forecast, where the $x$-axis represents the reserve bid 
size (MW) and the $y$-axis represents the shortfall probability 
$\varepsilon_t$. Three reliability thresholds are illustrated using dashed arrows: 
$1-\varepsilon_t = 0.99$ (blue), $1-\varepsilon_t = 0.90$ (orange, 
corresponding to the current P90 requirement), and $1-\varepsilon_t = 
0.80$ (green), each leading to a different optimal reserve bid size indicated by the value where the solid arrows meet the $x$-axis. The 
right plot shows the corresponding reserve merit-order curves. A stricter 
reliability threshold reduces the wind farm's bid size, shifts the 
merit-order curve leftward, and increases the reserve clearing price 
$\lambda_t$ (\euro/MW). The dashed line denotes reserve demand $D_t$ (MW).}}
\vspace{-0.5cm}
    \label{fig:bid_and_merit_schematic}
\end{figure}

Wind farms, EV aggregators, and dispatchable generators have historically 
been the dominant providers of FCR-D capacity in Denmark 
\cite{StatisticsAncillaryServices}, and the analysis therefore naturally 
centers on these three resource types. Within each technology class, we 
model a single representative agent with a joint capacity and pricing 
scheme. This representative-agent abstraction reflects the aggregated 
nature of reserve participation and preserves the key heterogeneity across 
resource types, while keeping the resulting bilevel model tractable. It 
also aligns with the fact that providers within a technology class share 
common operational characteristics, stochastic drivers, and regulatory 
obligations, even though intra-technology heterogeneity is not modeled 
explicitly.

We model the interaction between the TSO and reserve market participants 
as a bilevel optimization problem with a Stackelberg structure, as 
illustrated in Fig.~\ref{fig:infoex1}. The TSO acts as the leader by 
setting a reliability threshold $1-\varepsilon_t$ for each hour $t \in 
\mathcal{T}$, which defines the minimum required probability of delivery 
for stochastic reserve providers, while a representative wind farm, 
a representative EV aggregator, and a representative dispatchable 
generator act as followers. When the threshold is allowed to vary across 
hours, we refer to it as a \textit{dynamic} reliability threshold; when a 
single threshold is enforced uniformly across all hours, we refer to it as 
a \textit{static} reliability threshold, denoted $1-\varepsilon$. The 
reliability threshold $1-\varepsilon_t$ is communicated to the EV 
aggregator and the wind farm, each of which solves a 
chance-constrained optimization problem based on its probabilistic forecast 
to maximize its reserve capacity bid size $\hat{b}_{t,i}$, where $i \in 
\mathcal{I}$ indexes reserve providers. As illustrated in 
Fig.~\ref{fig:bid_and_merit_schematic} for the wind farm, a 
stricter reliability threshold reduces its bid size, shifts the reserve 
merit-order curve, and consequently increases the reserve market-clearing 
price $\lambda_t$. The same reasoning applies to the EV aggregator, whose 
bid size is likewise determined by the reliability threshold and its 
probabilistic forecast.

The representative dispatchable generator submits bids constrained only 
by its available capacity, as it is not subject to the reliability 
threshold. All bids are cleared in the reserve market-clearing problem 
given reserve demand $d_t$, yielding cleared capacities $b_{t,i}$ and 
an associated provision cost. The TSO anticipates these lower-level 
outcomes and minimizes total reserve-related cost, comprising reserve 
provision cost and expected shortfall penalties, by selecting the optimal 
reliability threshold.

\vspace{-5mm}
\subsection{Upper-level problem: TSO cost-minimization problem}
\nomenclature[D]{\(1-\varepsilon_t\)}{Threshold for reserve reliability at hour $t$ [-]}
\nomenclature[Da]{\(s_{t,i,\omega}\)}{Provider $i$'s reserve shortfall under scenario $\omega$ at hour $t$ [MW]}
\nomenclature[Cb]{\(\pi_{t,\omega}\)}{Probability of scenario $\omega$ at hour $t$ [-]}
\nomenclature[Cb]{\(\rho^{\rm sys}\)}{Penalty for reserve capacity shortfall compromising system security at hour $t$ [\euro/MW]}
\nomenclature[Cb]{\(\rho^{\rm viol}\)}{Penalty for reserve capacity shortfall at hour $t$ [\euro/MW]}
\nomenclature[Cb]{\(1-\varepsilon^{\rm sys}\)}{Threshold for system-level reserve reliability [-]}
\nomenclature[D]{\(d_t\)}{Cleared reserve demand at hour $t$ [MW]}
\nomenclature[Da]{\( s_{t,\omega}^{\rm viol}\)}{Total provider-level reserve shortfall under scenario $\omega$ at hour $t$ [MW]}
\nomenclature[Da]{\( s_{t,\omega}^{\rm sys}\)}{System-level reserve shortfall compromising system reliability under scenario $\omega$ at hour $t$ [MW]}
\nomenclature[Ce]{\( D_t \)}{Minimum reserve to be procured at hour $t$ [MW]}
\nomenclature[Db]{\( y_{t,\omega} \)}{Binary variable indicating system failure, where $y_{t,\omega}=1$ if a reserve capacity shortfall occurs under scenario $\omega$, and $y_{t,\omega}=0$ otherwise}
\nomenclature[A]{\( c_{t,i}(\cdot)\)}{Cost function of reserve provider $i$ at hour $t$}

The TSO minimizes its total reserve-related cost over decision variables 
including the reliability threshold $1-\boldsymbol{\varepsilon}\in
\mathbb{R}^{|\mathcal{T}|}$, the required reserve quantities 
$\boldsymbol{d}\in\mathbb{R}^{|\mathcal{T}|}$, the reserve shortfalls 
$\boldsymbol{s}\in\mathbb{R}^{|\mathcal{T}|\times|\mathcal{I}|
\times|\Omega|}$, and the system failure indicator $\boldsymbol{y}\in
\{0,1\}^{|\mathcal{T}|\times|\Omega|}$, where $\omega\in\Omega$ denotes 
scenarios of available reserve capacity observed by the TSO.

The objective in \eqref{eq:TSO objective} minimizes two cost components 
over a 24-hour horizon. The first is the reserve provision cost, 
determined by the cleared reserve capacities $b_{t,i}$ and the cost 
function $c_{t,i}(1-\varepsilon_t)$, detailed in Section 
\ref{cost_function}. The second is the expected reserve shortfall cost, 
comprising two penalty terms weighted by scenario probability 
$\pi_{t,\omega}$: a penalty $\rho^{\rm{viol}}\geq0$ on aggregate 
provider-level shortfalls, and a penalty $\rho^{\rm{sys}}\geq0$ on 
system-level shortfalls when total cleared capacity falls short of 
required demand $D_t$. Overbidding always carries a consequence, with a 
larger penalty applied when a provider shortfall additionally threatens 
system security. Both penalties are fixed parameters varied in the 
sensitivity analysis of Section \ref{sec:results}. Note that the 
objective is cast in terms of $c_{t,i}(1-\varepsilon_t)$ rather than 
the reserve market-clearing price $\lambda_t$, since using $\lambda_t$ 
would tie the cost solely to the marginal reserve provider, whose 
response to changes in the reliability level is not guaranteed, making 
the analysis less informative. Together, the two cost components capture 
the central trade-off: a high reliability threshold restricts stochastic 
providers to smaller bids, forcing the TSO to rely more heavily on 
expensive dispatchable generators, whereas a low threshold permits larger 
stochastic bids at the cost of greater shortfall risk and higher expected 
penalties:
\begin{subequations}\label{eq:TSO FULL OPT}
    \begin{align}
        \min_{\boldsymbol{\varepsilon},\boldsymbol{d},
        \boldsymbol{s},\boldsymbol{y}}\quad& 
        \overbrace{\sum_{t,i} c_{t,i}(1-\varepsilon_t)b_{t,i}}^
        \text{reserve provision cost} \nonumber \\ &\qquad+ 
        \overbrace{\sum_{t,\omega}\pi_{t,\omega}\left( \rho^{\rm{viol}}
        s_{t,\omega}^{\rm viol} + \rho^{\rm{sys}} 
        s_{t,\omega}^{\rm sys}\right)}^
        \text{reserve shortfall cost} . 
        \label{eq:TSO objective} \\
\intertext{\indent The objective penalizes two types of 
reserve shortfall. The first arises at the provider level: 
when the realized available reserve $r_{t,i,\omega}$ of 
provider $i$ under scenario $\omega$ falls short of the 
accepted bid $b_{t,i}$, an individual shortfall 
$s_{t,i,\omega}$ occurs:}
        &s_{t,i,\omega} \geq 0, &&\hspace{-2cm}
        \forall t,i,\omega, \label{eq:shortfall1} \\
        &s_{t,i,\omega} \geq b_{t,i} - r_{t,i,\omega},
        &&\hspace{-2cm}\forall t,i,\omega. 
        \label{eq:shortfall2} \\
\intertext{\indent The aggregate provider-level shortfall 
$s_{t,\omega}^{\rm viol}$, penalized by 
$\rho^{\rm{viol}}$ in \eqref{eq:TSO objective}, sums all 
individual provider shortfalls under scenario $\omega$:}
        & s_{t,\omega}^{\rm viol} = \sum_i s_{t,i,\omega}, 
        &&\hspace{-2cm}\forall t,\omega. 
        \label{eq:TSO system shortfall}\\
\intertext{\indent The second type of shortfall arises at 
the system level: when $s_{t,\omega}^{\rm viol}$ is large 
enough that total cleared capacity falls short of required 
demand $D_t$, system security is compromised. This shortfall $s_{t,\omega}^{\rm sys}$, 
penalized by $\rho^{\rm{sys}}$ in \eqref{eq:TSO objective}, 
is defined as:}
        &s^{\rm sys}_{t,\omega}\geq 0,&&\hspace{-2cm}
        \forall t,\omega, \label{eq: TSO compromising 
        system LL}\\
        &s^{\rm sys}_{t,\omega}\geq D_t-d_t+
        s_{t,\omega}^{\rm viol},&&\hspace{-2cm}\forall 
        t,\omega.\label{eq: TSO compromising system}\\
\intertext{\indent To limit the frequency of system-level 
shortfall events,  
\eqref{eq:TSO binary constraint}--\eqref{eq:TSO binary 
domain} enforce that the cleared reserve demand $d_t\geq 0$ 
satisfies reference demand $D_t$ with at least 
$1-\varepsilon^{\rm sys}$ reliability, where 
$\varepsilon^{\rm sys}$ is an exogenous parameter 
representing the maximum tolerated system-level shortfall 
probability. Unlike the provider-level threshold 
$1-\varepsilon_t$, which is optimized endogenously by the 
TSO, $1-\varepsilon^{\rm sys}$ is fixed exogenously and 
reflects a regulatory requirement on overall system 
reliability. This is enforced using Big-M notation with 
sufficiently large constant $M$, set to the minimum demand 
$D_t$, and binary variable $y_{t,\omega}\in\{0,1\}$ defined 
in \eqref{eq:TSO binary domain}: when $y_{t,\omega}=0$, 
\eqref{eq:TSO binary constraint} is binding and the 
scenario is deemed reliable; when $y_{t,\omega}=1$, the 
Big-M term relaxes the constraint, classifying the scenario 
as a system shortfall event. Constraint 
\eqref{eq: TSO shortfall sum of bins} ensures that the 
number of reliable scenarios meets 
$|\Omega|(1-\varepsilon^{\rm sys})$. One may disallow 
system shortfalls entirely by setting 
$1-\varepsilon^{\rm sys}=1$:}
        &d_t - s_{t,\omega}^{\rm viol} + My_{t,\omega} 
        \geq D_t, &&\hspace{-2cm}\forall t,\omega, 
        \label{eq:TSO binary constraint} \\
        &|\Omega|-\sum_\omega y_{t,\omega} \geq 
        |\Omega|(1-\varepsilon^{\rm sys}),&&\hspace{-2cm}
        \forall t, \label{eq: TSO shortfall sum of bins} \\
        &y_{t,\omega}\in\{0,1\},&&\hspace{-2cm}\forall 
        t,\omega. \label{eq:TSO binary domain} \\
\intertext{\indent Any reliability requirement below 80\% 
is assumed undesirable, reflecting a practical lower bound 
on acceptable participation standards:}
        &0.8\leq (1-\varepsilon_t) \leq 1,&&\hspace{-2cm}
        \forall t. \label{eq:TSO epsbounds}\\
\intertext{\indent In addition to its own objective and 
 \eqref{eq:TSO objective}--\eqref{eq:TSO 
epsbounds}, the TSO is subject to the optimal solutions of 
the lower-level problems:}
        &b_{t,i}\in\arg\min\{\text{lower-level problems 
        \eqref{eq:unit FULL}--\eqref{eq:market FULL}}\},
    \end{align}
\end{subequations}
which are stated in the following sections. In total, we have $(|\mathcal{I}|+1)|\mathcal{T}|$ lower-level problems, comprising $|\mathcal{I}|$ reserve provider problems and one reserve market-clearing problem per hour, over a horizon of $|\mathcal{T}|=24$ hours.

\subsection{Lower-level problem: Reserve providers}

\nomenclature[D]{\(\Hat{b}_{t,i}\)}{Reserve capacity bid offered by provider $i$ at hour $t$ [MW]}
\nomenclature[Cf]{\(R_{t,i}, r_{t,i,\omega}\)}{Random variable representing the reserve capacity availability of provider $i$ at hour $t$, and its realization under scenario $\omega$ [MW]}
\nomenclature[Ea]{\(\underline{\mu}_{t,i}, \Bar{\mu}_{t,i}\)}{Dual variables associated with the bounds of the reserve bid by provider $i$ at hour $t$ [MW]}

Each stochastic reserve provider $i\in\mathcal{I}$, represented by the representative EV aggregator and the representative wind farm in Fig.~\ref{fig:infoex1}, solves the following problem for each hour $t\in\mathcal{T}$. The objective in \eqref{eq:unit objective} is to maximize the reserve capacity bid size $\hat{b}_{t,i}$. This is equivalent to profit maximization: since providers are price takers, the market-clearing price $\lambda_t$ is a fixed revenue rate per MW of cleared capacity, and FCR-D involves no activation-related costs, as discussed in Section~\ref{prelim}. The chance constraint \eqref{eq:unit CC} enforces that the bid does not exceed the available reserve capacity $R_{t,i}$ with at least $1-\varepsilon_t$ probability, where $R_{t,i}$ is a continuous random variable determined by the provider's probabilistic forecast, with realizations $r_{t,i,\omega}$. Constraint \eqref{eq:unit LL} declares non-negativity of the reserve bids:
\begin{subequations}\label{eq:unit FULL}
    \begin{align}
        \max_{\Hat{b}_{t,i}} \quad &\Hat{b}_{t,i} 
        \label{eq:unit objective} \\
        \text{s.t.}\quad&\mathbb{P}\left(\Hat{b}_{t,i} 
        \leq R_{t,i}\right) \geq 1-\varepsilon_t, 
        &(\Bar{\mu}_{t,i}) \label{eq:unit CC}\\
        &-\Hat{b}_{t,i} \leq 0, 
        &(\underline{\mu}_{t,i}), \label{eq:unit LL}
    \end{align}
\end{subequations}
where the reliability threshold $1-\varepsilon_t$ in \eqref{eq:unit CC} is treated as a fixed input parameter, since stochastic providers take the TSO's threshold as given and optimize their bids accordingly. The chance constraint \eqref{eq:unit CC} is subsequently reformulated analytically into a tractable deterministic constraint in Section~\ref{chance1}, enabling embedding into a single-level reformulation.

The representative dispatchable generator is not subject to the reliability threshold; its bid $\hat{b}_{t,i}$ is bounded between zero and its available capacity.

\subsection{Lower-level problem: Reserve market-clearing}
\nomenclature[D]{\(b_{t,i}\)}{Cleared reserve capacity of provider $i$ at hour $t$ [MW]}
\nomenclature[Ca]{\(\alpha_{t,i}\),\(\beta_{t,i}\)}{Intercept and slope coefficients of the cost function for reserve provider $i$ at hour $t$ [\euro/MW]}
\nomenclature[Eb]{\(\underline{\nu}_{t,i}, \Bar{\nu}_{t,i}\)}{Dual variables associated with the bounds on the cleared reserve capacity of provider $i$ at hour $t$ [\euro/MW]}
\nomenclature[Ec]{\(\lambda_t\)}{Reserve market-clearing price at hour $t$ [\euro/MW]}

The reserve market-clearing problem, represented by the bottom-right node in Fig.~\ref{fig:infoex1} and illustrated in detail in Fig.~\ref{fig:bid_and_merit_schematic}, clears bids from all reserve providers on an hourly basis for each $t\in\mathcal{T}$. The objective function in \eqref{subeq:marketobj} minimizes the total reserve procurement cost, where the cost function $c_{t,i}(1-\varepsilon_t)$ is identical to that in \eqref{eq:TSO objective} and is formally defined later in Section~\ref{cost_function}. The objective can be interpreted as assuming that each provider submits its reserve bid at a price equal to the reliability-driven cost $c_{t,i}(1-\varepsilon_t)$, rather than at a bid price reflecting potential opportunity costs associated with reserve capacity booking. This assumption is reasonable in the context of the FCR-D reserve product considered in this paper, as FCR-D is generally not energy intensive. Hence, the opportunity cost of reserving capacity is expected to be limited and can be neglected. Constraints \eqref{subeq:supply bounds market clearing} and \eqref{subeq:market LB} enforce the cleared reserve capacity $b_{t,i}$ of each provider to lie between zero and the submitted bid $\hat{b}_{t,i}$. Since  \eqref{subeq:supply bounds market clearing} is conditioned on the optimal solution of \eqref{eq:unit FULL}, it falls into the class of generalized Nash constraints \cite{HARKER199181}. The market-clearing constraint \eqref{subeq:supply and demand} ensures that the total cleared reserve capacity satisfies the reserve demand $d_t$ specified by the TSO:

\begin{subequations} \label{eq:market FULL}
    \begin{align}
        \min_{b_{t,i}} &\quad \sum_i c_{t,i}(1-\varepsilon_t)b_{t,i} \label{subeq:marketobj} \\
        \text{s.t.}&\quad b_{t,i} \leq \Hat{b}_{t,i}, \quad \forall i, & (\overline{\nu}_{t,i}) \label{subeq:supply bounds market clearing} \\
        &\quad -b_{t,i}\leq 0, \quad \forall i, & (\underline{\nu}_{t,i}) \label{subeq:market LB} \\
        &\quad d_{t} - \sum_i b_{t,i} = 0,& (\lambda_t). \label{subeq:supply and demand}
    \end{align}
\end{subequations}

Since the reserve demand $d_t$ is set by the TSO in the 
upper level, it enters the market-clearing problem as a 
fixed parameter. Similarly, the bid bounds $\hat{b}_{t,i}$ 
are determined by the reserve providers in 
\eqref{eq:unit FULL}, and the cost function 
$c_{t,i}(1-\varepsilon_t)$ depends on the reliability 
threshold set by the TSO. Consequently, all inputs to the 
market-clearing problem are fixed at this level, making it 
a linear program whose optimal solution yields the cleared 
capacities $b_{t,i}$ and the reserve market-clearing price 
$\lambda_t$, the latter being the dual variable associated 
with \eqref{subeq:supply and demand}. A discussion on the uniqueness of the optimal solution in the market-clearing given the TSO's decision $1-\varepsilon^*_t$ and the reserve providers' response $\hat{b}_{t,i}^*$ can be found in Appendix \ref{app:uniqueness stackelberg}.

\label{cost_function}
\subsection{Cost function definition}

The provision cost of stochastic reserve providers is assumed to be 
increasing in the reliability threshold set by the TSO. Specifically, 
a higher reliability threshold requires the provider to ensure 
availability with greater probability, reducing the feasible bid size 
and warranting a higher unit cost as compensation. Conversely, a lower 
threshold allows larger bids, with a reduced price reflecting the 
transfer of risk to the TSO. We therefore define the cost function of 
reserve provider $i$ at hour $t$ as:
\begin{equation} \label{eq:cost function}
    c_{t,i}(1-\varepsilon_t)=\alpha_{t,i} + 
    (1-\varepsilon_t)\beta_{t,i},
\end{equation}
where $\alpha_{t,i}\geq 0$ is the intercept and $\beta_{t,i}\geq 0$ 
is the reliability-dependent slope, both specific to provider $i$ and 
hour $t$.\footnote{This linear functional form is motivated in part 
by discussions with Energinet, whose operational experience suggests 
a monotone increasing relationship between required reliability and 
provider compensation. Exact bid data are commercially confidential 
and not publicly available.} The cost function is increasing in the 
reliability threshold: as $1-\varepsilon_t\rightarrow 1$, the bid 
size shrinks and the unit cost rises, while as 
$1-\varepsilon_t\rightarrow 0$, larger bids are permitted at a lower 
unit cost. For dispatchable generators, the provision cost is 
independent of the reliability threshold, corresponding to 
$\beta_{t,i}=0$. This linear specification is a natural first-order 
approximation, consistent with the FCR-D market structure where 
providers are compensated solely through a capacity payment and 
activated energy is settled through the imbalance mechanism.

\label{chance1}
\subsection{Analytical reformulation of chance constraint}

Problem \eqref{eq:unit FULL} involves a chance constraint that could in principle be handled using sample-based approaches \cite{NemirovskiShapiro06}. However, because the chance constraint appears in the lower level, such methods would lead to a substantial increase in variables and constraints when deriving the optimality conditions, particularly at high reliability levels \cite{SampleSizeSAA}. We therefore adopt an analytical reformulation, which avoids these scalability issues and makes the problem computationally tractable, at the cost of imposing a distributional assumption on the forecast tail discussed below.

Since the reliability threshold in \eqref{eq:TSO epsbounds} varies between 80\% and 100\%, we restrict attention to the relevant tail of the forecast distribution rather than fitting the entire distribution. Any bid size beyond the 20\textsuperscript{th} percentile would correspond to a reliability level below 80\% and is therefore infeasible. We denote the available reserve at the 20\textsuperscript{th} percentile by $r_{t,i}^{0.2}$, defined such that $\mathbb{P}(R_{t,i}\leq r_{t,i}^{0.2})=0.2$, where $R_{t,i}$ is a continuous random variable representing the probabilistic forecast of provider $i$ at hour $t$. This differs from \cite{TRH}, where the distribution is fitted below the 10\textsuperscript{th} percentile.

Applying Bayes' theorem to express the conditional 
probability of the realized reserve $r_{t,i,\omega}$ 
falling below the bid $\hat{b}_{t,i}$, the chance 
constraint \eqref{eq:unit CC} is reformulated as follows, 
with intermediate calculations provided in 
Appendix~\ref{app:bayes etc}:
\begin{subequations}
    \begin{align}
        &\mathbb{P}\left(R_{t,i}\leq\hat{b}_{t,i}\mid 
        R_{t,i}\leq r_{t,i}^{0.2}\right)\leq
        \frac{\varepsilon_t}{0.2}. \\
\intertext{Denoting the conditional CDF as $F(\hat{b}_{t,i})$, this becomes:}
        &F(\hat{b}_{t,i})\leq\frac{\varepsilon_t}{0.2}. \\
\intertext{Assuming existence of the inverse CDF yields a deterministic 
upper bound on the bid:}
        &\hat{b}_{t,i}\leq F^{-1}
        \left(\frac{\varepsilon_t}{0.2}\right). \label{eq:reformulated CC} \\
\intertext{For notational convenience we define:}
        &F^{-1}\left(\frac{\varepsilon_t}{0.2}\right)
        =:z_{t,i}. \label{eq:unit inverse CDF}
            \end{align}
\end{subequations}

\nomenclature[A]{\(F(\cdot)\)}{Cumulative distribution 
function of available reserve}

We model the tail of the forecast distribution using the 
two-parameter Weibull distribution, chosen for its flexible 
behavior, and crucially its closed-form inverse, which is 
essential for the analytical reformulation \cite{Weibull}. The CDF is:
\begin{equation}
    F(\Hat{b}_{t,i})=\begin{cases}1-\exp\left(-\kappa (\Hat{b}_{t,i})^{\gamma}\right),&\Hat{b}_{t,i}\geq 0,\\0,&\Hat{b}_{t,i}<0,
    \end{cases} \label{eq:WeibullCDF}
\end{equation} 
where $\kappa>0$ is the scale parameter and $\gamma>0$ is 
the shape parameter. The chance constraint \eqref{eq:unit 
CC} therefore reduces to the deterministic constraint 
$\hat{b}_{t,i}\leq z_{t,i}$, which is linear in 
$\hat{b}_{t,i}$ for a fixed threshold $\varepsilon_t$.

For $\gamma \leq 1$, \eqref{eq:WeibullCDF} is globally concave, whereas for $\gamma > 1$ it exhibits an inflection point where it transitions from convex to concave, as shown in Appendix~\ref{app:convexity of CDF}. The convexity properties of the inverse function are discussed in Appendix~\ref{app:convexity of inverse CDF}. Note that the fitted distributions in our analysis exhibit inflection points for most hours, yet this does not affect the optimal solution, which is discussed in Appendix~\ref{app:uniqueness stackelberg} along with the uniqueness of the reserve providers' optimal solution given the TSO's decision $1-\varepsilon_t$.

\subsection{Single-level optimization problem}
The bilevel optimization problem is reformulated as a single-level problem by deriving the optimality conditions of the lower-level problems \eqref{eq:unit FULL} and \eqref{eq:market FULL} and appending them as constraints to the upper level \cite{Gabriel}. The resulting formulation in \eqref{eq:FULL OPT} includes binary variables from \eqref{eq:TSO binary constraint} and \eqref{eq: TSO shortfall sum of bins}, the nonlinear function in \eqref{eq:unit inverse CDF}, and bilinear terms in \eqref{eq:strong duality1} and \eqref{eq:strong duality2}. Consequently, the resulting model is a mixed-integer nonlinear optimization problem, which we subsequently linearize to obtain a computationally efficient formulation.

Rather than enforcing complementarity conditions for each lower-level problem directly, which would introduce multiple nonlinear products, one could impose a single strong duality condition requiring equality between the primal and dual objective functions at optimality. Instead, we impose the \textit{reverse weak duality} conditions in \eqref{eq:strong duality1} and \eqref{eq:strong duality2}, requiring the dual objective (maximization) to be greater than or equal to the primal objective (minimization). This does not affect the optimal solution, since primal and dual objective values must coincide at optimality. However, we find it is computationally advantageous because it relaxes the feasible region relative to enforcing equality directly.

With the set of decision variables $\boldsymbol{\xi}\in\{
\varepsilon_t$, $d_t$, $s_{t,i,\omega}$, $s_{t,\omega}^{\rm sys}$,
$s_{t,\omega}^{\rm viol}, y_{t,\omega}, z_{t,i}, 
\hat{b}_{t,i}, \bar{\mu}_{t,i}, \underline{\mu}_{t,i}, 
b_{t,i}, \bar{\nu}_{t,i}, \underline{\nu}_{t,i}, 
\lambda_t\}$, $\forall t,i,\omega$, the resulting single-level optimization 
problem is \eqref{obb}-\eqref{last}:
\begin{subequations}\label{eq:FULL OPT}
    \begin{align}
        \min_{\boldsymbol{\xi}}\quad&
        \eqref{eq:TSO objective} \label{obb}\\
        \text{s.t.}\quad&
        \eqref{eq:shortfall1}-\eqref{eq:TSO epsbounds}, 
\intertext{with the primal feasibility, stationarity, 
and reverse weak duality from \eqref{eq:unit FULL}, 
applying the analytical reformulation 
\eqref{eq:unit inverse CDF}:}
        &\eqref{eq:unit LL}, \eqref{eq:reformulated CC},
        \\
        &-1+\bar{\mu}_{t,i}-\underline{\mu}_{t,i}=0, 
        &&\hspace{-1cm}\forall t,i, \label{eq:stationarity1}\\
        &-\bar{\mu}_{t,i}z_{t,i}\geq-\hat{b}_{t,i}, 
        &&\hspace{-1cm}\forall t,i, \label{eq:strong duality1}\\
\intertext{with the primal feasibility, stationarity, and 
reverse weak duality from \eqref{eq:market FULL}, applying the cost function \eqref{eq:cost function}:}
        &\eqref{subeq:supply bounds market clearing}-\eqref{subeq:supply and demand},
        \\
        &\alpha_{t,i} + (1-\varepsilon_t)\beta_{t,i}-\underline{\nu}_{t,i}+
        \bar{\nu}_{t,i}-\lambda_t=0, &&\forall t,i, 
        \label{eq:stationarity2}\\
        &\sum_i-\bar{\nu}_{t,i}\hat{b}_{t,i}+
        \lambda_t d_t \nonumber\\&\hspace{1.5cm}\geq\sum_i \left(\alpha_{t,i} + (1-\varepsilon_t)\beta_{t,i}\right)
        b_{t,i}, &&\forall t, \label{eq:strong duality2}\\
\intertext{and with dual feasibility from \eqref{eq:unit FULL} and \eqref{eq:market FULL}:}
        &\bar{\mu}_{t,i},\underline{\mu}_{t,i},
        \bar{\nu}_{t,i},\underline{\nu}_{t,i}\geq 0, 
        \quad\lambda_t\text{ free}, &&\hspace{-1cm}\forall t,i. \label{last}
    \end{align}
\end{subequations}

The single-level formulation \eqref{eq:FULL OPT} is a mixed-integer nonlinear problem, and contains both binary and nonlinear components. The binary variables $y_{t,\omega}$ arise from the shortfall penalty modeling in \eqref{eq:TSO binary constraint} and \eqref{eq: TSO shortfall sum of bins}. The nonlinear terms consist of the bilinear products $\varepsilon_t b_{t,i}$ in \eqref{eq:TSO objective}, $\bar{\mu}_{t,i}z_{t,i}$ in \eqref{eq:strong duality1}, and $\bar{\nu}_{t,i}\hat{b}_{t,i}$, $\lambda_t d_t$, and $\varepsilon_t b_{t,i}$ in \eqref{eq:strong duality2}, as well as the nonlinear inverse CDF $F^{-1}(\varepsilon_t/0.2)$ in \eqref{eq:unit inverse CDF}. To obtain a computationally tractable mixed-integer \textit{linear} formulation, all nonlinear terms are linearized using McCormick envelopes, as detailed in Appendices~\ref{app:mccormick} and \ref{app:dual upper bounds}, and piecewise-linear approximations, as detailed in Appendix~\ref{app:linearizations of f}. The complete optimization program is available in the companion repository \cite{Git}.

\section{Case Study and Results}\label{sec:case and validation methods}

The case study is grounded in the Danish FCR-D up-regulation market,  where wind farms, EV aggregators, and  conventional generators are the  three historically dominant provider types \cite{StatisticsAncillaryServices}.  The following describes how each representative provider is  parameterized using real Danish data.

We consider historical flexibility data from an EV fleet representing  the EV aggregator participating in the reserve market. The dataset consists of charging profiles from 1,400 residential EV charging boxes  in Denmark, recorded between March 24, 2022, and March 21, 2023, with one EV assumed to be coupled to each charging box. The historical EV consumption level serves as a baseline for flexibility estimation, where EVs may offer up-regulation by reducing their charging rate. The available flexibility for up-regulation is provided on an hourly basis; for details on how it is estimated, the reader is referred to \cite{TRH}.

To represent the wind farm participating in the reserve market, we assume that the producer can adjust output through blade pitching, enabling both increases and decreases in momentary production. Hourly 
day-ahead wind speed forecasts at a given location are simulated using CorRES \cite{CorRES}, which accounts for both the lag-time and forecast  horizon at the time of simulation. These wind speeds are then translated  into power forecasts using the power curve and rated capacity of the  relevant wind farm. Based on these forecasts, we assume the wind farm  can increase production by 5\% to provide up-regulation, meaning that  5\% of the forecasted power output is used as the available reserve 
capacity in our analysis.

Wind forecast uncertainty has historically been modeled using a variety  of distributions, including the two-parameter Weibull and Beta  distributions \cite{CARTA2009933}. However, when a single distribution  is fitted to the entire sample pool, the tails of the data are often  poorly captured, which can lead to insufficient reserve bids from the  wind producer at high reliability levels. To address this, we apply the  methodology described above, fitting a two-parameter Weibull 
distribution to samples below the 20th percentile using  \eqref{eq:WeibullCDF}.

\begin{figure}[t]
    \centering
    \includegraphics[scale=1]{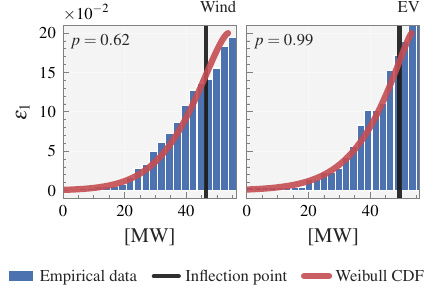}
    \caption{\small{Cumulative histogram of empirical samples below the 20th percentile with fitted Weibull cumulative distribution function and associated $p$-values following \cite{KSTest} of wind forecast with $p=0.62$ (left) and EV flexibility with $p=0.99$ (right) for hour 1. The $p$-values indicate that these are sufficient fits to represent the tails of the sample pool. The inflection point indicates where the distribution changes from convex to concave.}}
    \label{fig:fitted distributions}
    \vspace{-4mm}
\end{figure}

Figure \ref{fig:fitted distributions} illustrates Weibull distributions  fitted to samples below the 20th percentile for a wind forecast and the  EV historical flexibility data. For the wind distribution, the instance  shown corresponds to the hour with the poorest fit across the 24  simulated hours, as determined by the Kolmogorov-Smirnov test  \cite{KSTest}, i.e., the hour with the lowest $p$-value. Since the fit  is only rejected for $p<0.05$ and this worst-case hour remains above  that threshold, all 24 hourly fits for all stochastic reserve providers  are accepted for further analysis. The black line marks the inflection  point of the CDF (the transition from convex to concave behavior) as 
discussed previously.

\subsection{Setup for the out-of-sample validation}
%\nomenclature[A]{\(\boldsymbol{1}_{x>0}(x) \)}{Indicator function, takes value of $1$ when $x>0$, and 0 otherwise}

The TSO quantifies provider-level reserve shortfalls in \eqref{eq:shortfall1}-\eqref{eq:shortfall2} to evaluate when additional reserves beyond the minimum amount are required. The available samples not used for distribution fitting and optimization are reserved for out-of-sample evaluation, allowing us to assess the performance of our model. For a sample $\omega'\in\Omega' \setminus \Omega$ in the out-of-sample set, a shortfall $s_{t,i,\omega'}$ occurs when the realization $r_{t,i,\omega'}$ falls short of the accepted bid $b_{t,i}$, constituting an out-of-sample overbid by provider $i$ in scenario $\omega'$. We employ two distinct reliability measures. First, the relative \textit{number} of shortfalls across the scenario set must not exceed the allowed threshold:
\begin{subequations}
    \begin{align}
        &\frac{1}{|\Omega'|}\sum_{\omega'\in\Omega' : s_{t,i,\omega'} > 0} \boldsymbol{1}_{s_{t,i,\omega'}>0}(s_{t,i,\omega'}) < \varepsilon_t, \ \ \ \forall t,i, \label{eq:shortfall percentage by number}
\intertext{where $\boldsymbol{1}_{s_{t,i,\omega'}>0}(s_{t,i,\omega'})$ is the indicator function; a mathematical function that describes the membership of an element in a specific subset. It maps elements inside the subset to 1 and elements outside to 0, in our case 1 when there is an out-of-sample shortfall and 0 otherwise. Second, for all scenarios in which a shortfall occurs, we measure the ratio of the shortfall to the accepted bid and average this across scenarios, yielding a reliability measure in terms of \textit{quantity}:}
        &\qquad\frac{1}{\vert \Omega'\vert}\sum_{\omega'\in \Omega' : s_{t,i,\omega'} > 0} \frac{s_{t,i,\omega'}}{b_{t,i}} < \varepsilon_t, \ \ \ \ \forall t,i. \label{eq:shortfall percentage by quantity}
    \end{align}
\end{subequations}

\subsection{Results: Optimal reliability threshold}\label{sec:results}
The mixed-integer linear version of \eqref{eq:FULL OPT} is solved under both dynamic and static reliability levels\footnote{Implemented using Gurobi v.13 and Python v.3.10 on an HP EliteBook 840 14-inch G10 Notebook PC (Windows 11), equipped with a 13th Gen Intel® Core™ i7-1365U processor (10 cores, 12 logical processors), using Visual Studio Code. The problem has 6,793 continuous and 1,200 binary variables.}, where dynamic (time-variant) reliability levels may vary across hours of the day, while static (time-invariant) levels apply a uniform threshold to all hours. All source codes are available in the companion repository \cite{Git}. Results are compared to Energinet's ad hoc P90 reliability requirement.

In the dynamic case, \eqref{eq:FULL OPT} solves in 1 second with a 0.0\% optimality gap, whereas the static case requires 15 seconds at the same gap; a slight increase in duration due to the added complexity of optimizing a single shared threshold across all hours simultaneously.

\begin{figure}[t]
\centering
\includegraphics[scale=1]{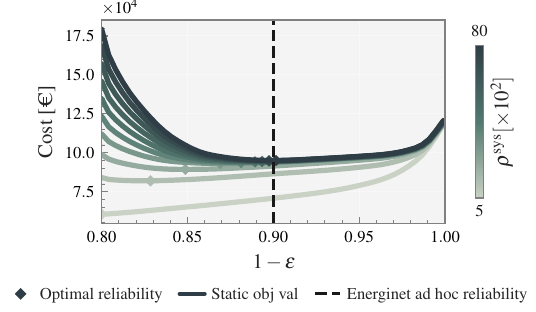}
\caption{\small{Pareto frontier for a static, time-invariant reliability threshold,  shown for different penalty values of $\rho^{\rm sys}$ with $\rho^{\rm viol} =  0.5\rho^{\rm sys}$. Each curve shows the optimal value of the objective  function \eqref{eq:TSO objective} as a function of the static threshold  $1-\varepsilon$. The diamond marks the cost-minimizing reliability threshold, compared against Energinet's ad hoc P90 requirement.  Depending on the penalty parameterization, endogenously determining  the threshold reduces total reserve-related costs by up to 14.47\% relative to the fixed P90 standard.}}
\vspace{-4mm}
\label{fig:pareto different penalty}
\end{figure}

Figure \ref{fig:pareto different penalty} summarizes the system-wide trade-off between cost and reliability under the static reliability threshold for different values of $\rho^{\rm sys}$ and $\rho^{\rm viol}=0.5\rho^{\rm sys}$, indicated by the color-bar. When a single threshold is applied uniformly across all hours, the Pareto front exhibits the expected convex shape; costs are elevated at both extremes of the reliability range but heavily depend on the penalty price. The minimum-cost solution occurs near $1-\varepsilon^*\in (0.8, 0.9)$, beyond which costs rise as the threshold approaches 100\%. This increase reflects the disproportionate cost of the final increments in reliability, where near-perfect reliability must be secured by expensive conventional dispatchable generation, yielding diminishing economic returns. On the other side, the costs at low reliability levels on the left side of the curves reflect the magnitude of the penalty price. A low penalty incentivizes setting the reliability threshold to a lower value and procuring reserve from stochastic providers, while a high penalty removes that incentive and rather sets the reliability threshold high.

When comparing the optimal cost to the P90 requirement, a cost  reduction of up to 14.47\% is possible by endogenously determining  the reliability threshold, depending on the shortfall penalty  parameterization. This indicates that Energinet's P90 standard is a  well-motivated starting point that approaches the optimal threshold  in some penalty regimes, but leaves room for cost improvement in  others.

In the case of dynamic reliability levels, the cost reduction ranges  between 0.6\% and 2.4\% relative to the optimal static threshold.  While modest in the context of a single representative wind farm, a  single representative EV aggregator, and a single representative  dispatchable conventional generator, these savings are expected to scale with the number of market participants, the volume of procured reserves, and the degree of temporal variability in the generation mix. In a real-world setting with a larger and more diverse pool of reserve providers, the efficiency gains from temporally differentiated reliability thresholds could be considerably more pronounced.

\begin{figure}[t]
    \centering
    \includegraphics[scale=1]{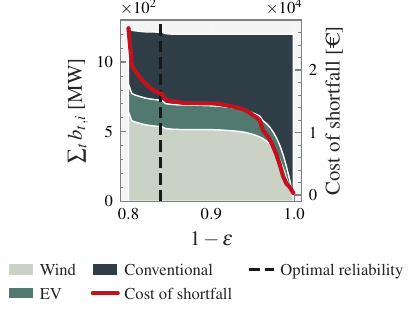}
    \caption{\small{Total stacked reserve provision mix as a function of the reliability  threshold $1-\varepsilon$, with the associated total shortfall cost shown in red.}}
    \label{fig:mix results}
    \vspace{-4mm}
\end{figure}

\begin{table}[b]
    \caption{\small{The percentage of reserve coming from each provider, and the associated bid price for an arbitrarily selected hour.}}
    \centering
    \begin{tabular}{|c|c|c|c|c|}\hline
    $1-\varepsilon$ & Wind (\euro/MW) & EV (\euro/MW) & Conventional (\euro/MW) \\ \hline
    0.80 & 0.79 (10.00) & 0.14 (20.00) & 0.07 (100.00) \\
    $0.84^*$ & 0.73 (13.84) & 0.12 (24.80) & 0.14 (100.00) \\
    1.00 & 0.00 (30.00) & 0.00 (45.00) & 1.00 (100.00) \\ \hline
    \end{tabular}
    \label{tab:merit order kind of}
    \vspace{-4mm}
\end{table}

\subsection{Results: Optimal reserve provision mix}
For a given penalty parameterization, Fig.~\ref{fig:mix results}  illustrates the aggregate reserve provision mix over the entire time  horizon across the range of possible reliability thresholds. For low 
reliability levels, the cheaper stochastic reserve providers contribute  a larger share of the total reserve provision mix, and the TSO clears  a reserve demand above the minimum requirement $D_t$ to buffer against 
the higher expected shortfalls associated with less reliable providers.  The point of lowest objective function value, and thus the optimal  reliability threshold, is at $1-\varepsilon^*=0.84$, beyond which the  reserve quantities from stochastic providers decrease along with the  cost of shortfalls. At high reliability levels, stochastic reserve  providers are fully excluded from the market and the system relies  entirely on the representative dispatchable conventional generator. This is  illustrated in Table~\ref{tab:merit order kind of}, which shows the  share of cleared reserve capacity from each representative provider  and the associated bid price at three reliability levels: $1-\varepsilon  = 0.8$, the optimal static threshold $1-\varepsilon^*$, and $1- \varepsilon = 1$. At low reliability levels, reserve provision is met  predominantly by the representative wind farm and the representative  EV aggregator at low marginal cost. As the reliability threshold  tightens toward 1, the reserve provision mix shifts toward the  representative dispatchable generator, reflecting a move to costlier  but fully reliable capacity. At the optimum $1-\varepsilon^*$, the  system strikes a balance between reserve adequacy and procurement  cost, at the point where the marginal cost of additional reliability  begins to rise steeply, consistent with the Pareto frontier behavior  discussed previously.

\subsection{Results: Out-of-sample validation \& sensitivity analysis}
\begin{figure}[t]
    \centering
    \includegraphics[scale=1]{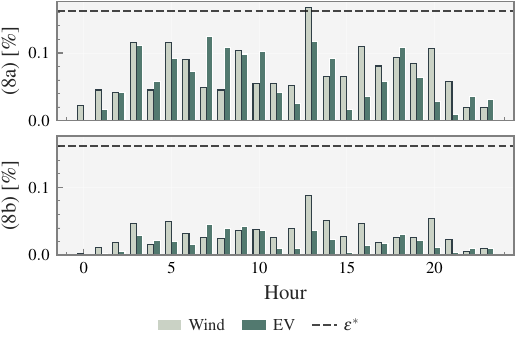}
    \caption{\small{Out-of-sample shortfall percentage measured by number of violations  in the top panel and by shortfall quantity in the bottom panel,  corresponding to \eqref{eq:shortfall percentage by number} and  \eqref{eq:shortfall percentage by quantity} respectively, for the  optimal static reliability threshold $\varepsilon^*$ with penalty  parameters $\rho^{\rm sys}=1000$, $\rho^{\rm viol}=0.5\rho^{\rm sys}$,  and system reliability requirement $1-\varepsilon^{\rm sys}=0.9$.}}
    \label{fig:shortfall dynamic}
    \vspace{-4mm}
\end{figure}

Fig.~\ref{fig:shortfall dynamic} shows the out-of-sample shortfall  percentages for the static reliability level, measured by number of  violations in the top panel and by shortfall quantity in the bottom  panel, corresponding to \eqref{eq:shortfall percentage by number} and  \eqref{eq:shortfall percentage by quantity} respectively. The two  panels highlight how the choice of quantification method materially  affects the assessment of reliability compliance: although shortfalls  are observed in all hours, the measured percentage is higher under the  number-based measure than under the quantity-based measure, indicating  that while violations occur, their magnitude is small. Similar results for the dynamic case are included in the companion repository \cite{Git}.

Table~\ref{tab:sensitivity analysis eps_sys} presents a sensitivity analysis of the TSO's total costs across varying system reliability levels for both the static and dynamic cases, with the resulting  optimal reliability thresholds shown in the table. Both the cost and  the resulting optimal thresholds are consistent throughout the range 
of possible system reliability levels below full coverage, indicating  that the optimal solutions do not strongly depend on $1-\varepsilon^{\rm sys}$ unless all shortfalls must be covered (i.e., $1-\varepsilon^{\rm sys}=1$). In that case, costs increase dramatically and the  optimal static threshold rises accordingly. In the dynamic case, the  total cost is systematically lower than in the static case, with  consistent magnitude across all system reliability levels. The mean  dynamic threshold remains close to the optimal static threshold, even  under perfect system reliability, suggesting that dynamic thresholds  absorb costly shortfall hours by selectively increasing reliability  in those specific hours rather than raising the threshold uniformly.

\section{Conclusions and future work}\label{sec:conclusions}
This paper treated the reliability threshold in reserve  markets as a design variable rather than a regulatory constant,  developing a bilevel optimization framework in which the TSO  determines the threshold endogenously by balancing reserve  procurement costs against expected shortfall penalties. Applied to  the Nordic FCR-D market and motivated by Energinet's P90 requirement,  the framework reveals that static regulatory thresholds result in  suboptimal procurement costs: endogenously determined static  thresholds reduce costs by up to 14.5\% relative to P90, and dynamic  hourly thresholds yield a further 2.4\%. The optimal threshold  consistently falls below P90, in the range $1-\varepsilon^*\in(0.8, 0.9)$, because near-perfect reliability forces excessive reliance on  expensive dispatchable generation while largely excluding cheaper  stochastic providers. The sensitivity analysis confirms this finding  is robust across a wide range of system reliability requirements, with  costs rising sharply only when zero shortfall tolerance is imposed.  Taken together, the results suggest that moving from a fixed  regulatory threshold to an endogenously optimized one is both  tractable via the single-level reformulation developed here and  economically consequential, with gains likely to be more pronounced  in larger and more diverse markets than the one studied here.

Several directions merit future investigation. Extending the framework  to energy-intensive products such as aFRR and mFRR would require  incorporating rebound effects and multi-period energy constraints.  Relaxing the representative-agent structure to accommodate a  heterogeneous pool of individually parameterized providers would  improve quantitative generalizability and capture richer operational  and stochastic diversity. The interaction between the reliability  threshold and voluntary market participation also deserves attention:  as the threshold tightens, marginal stochastic providers may exit  entirely, an extensive-margin response the current model does not capture. Finally, extending the framework to a multi-service setting in which providers simultaneously participate in energy and multiple reserve markets would make the opportunity cost of reserve capacity endogenous to the optimization.

\begin{table}
    \caption{\small{Sensitivity on the total cost with different values of $1-\varepsilon^{\rm sys}$, with $\rho^{\rm sys}=1000$ and $\rho^{\rm viol}=0.5\rho^{\rm sys}$.}}
    \centering
    \begin{tabular}{|c|c|c|c|c|}\hline
    &\multicolumn{2}{c|}{Static} & \multicolumn{2}{c|}{Dynamic} \\ \hline
    $1-\varepsilon^{\rm sys}$ & \eqref{eq:TSO objective} & $1-\varepsilon^*$ & \eqref{eq:TSO objective} & $\overline{1-\varepsilon}^*_t (\sigma_{1-\varepsilon^*_t})$ \\ \hline
    0.7 & 77471.27 & 0.84 & 76562.81 & 0.84 (0.04) \\
    0.8 & 77465.63 & 0.84 & 76102.33 & 0.83 (0.03) \\
    0.9 & 77480.28 & 0.84 & 76410.31 & 0.83 (0.04) \\
    1.0 & 91303.50 & 0.88 & 88071.37 & 0.85 (0.03) \\ \hline
    \end{tabular}
    \label{tab:sensitivity analysis eps_sys}
       \vspace{-4mm}
\end{table}

\appendix
\subsection{Reformulation of chance constraint calculations}\label{app:bayes etc}
Denoting the available reserve at the 20\textsuperscript{th} percentile by $r_{t,i}^{0.2}$, i.e., $\mathbb{P}(R_{t,i} \leq r_{t,i}^{0.2})=0.2$, where, as before, $R_{t,i}$ is a continuous random variable following a distribution depending on the forecast of the individual reserve provider $i$ at hour $t$, we express the conditional probability of a realization $r_{t,i,\omega}$ of the random variable $R_{t,i}$ being under the reserve providers placed bid $\Hat{b}_{t,i}$ using Bayes' theorem:
\begin{subequations}
    \begin{align}
        &\mathbb{P}\left(R_{t,i} \leq \Hat{b}_{t,i} \mid R_{t,i} \leq r_{t,i}^{0.2}\right)\nonumber \\ &= \frac{\mathbb{P}\left(R_{t,i} \leq \Hat{b}_{t,i}\right)\mathbb{P}\left(R_{t,i} \leq r_{t,i}^{0.2} \mid R_{t,i} \leq \Hat{b}_{t,i}\right)}{\mathbb{P}\left(R_{t,i} \leq r_{t,i}^{0.2}\right)} \\
        \Rightarrow& \mathbb{P}\left(R_{t,i} \leq \Hat{b}_{t,i}\right)= 0.2\mathbb{P}\left(R_{t,i} \leq \Hat{b}_{t,i} \mid R_{t,i} \leq r_{t,i}^{0.2}\right), \label{eq:Bayes theorem}
\intertext{where it is implicit that $\mathbb{P}\left(R_{t,i} \leq r_{t,i}^{0.2} \mid R_{t,i} \leq \Hat{b}_{t,i}\right) = 1$, and $\mathbb{P}(R_{t,i} \leq r_{t,i}^{0.2})=0.2$ as previously stated. We then have:}
        &1-\mathbb{P}\left(R_{t,i}\leq\Hat{b}_{t,i} \right) \geq 1 - \varepsilon_t, \\
\intertext{which by \eqref{eq:Bayes theorem} and some basic arithmetic is the same as:}
        &0.2\mathbb{P}\left(R_{t,i} \leq \Hat{b}_{t,i} \mid R_{t,i} \leq r_{t,i}^{0.2}\right) \leq \varepsilon_t.
    \end{align}
\end{subequations}

\subsection{Convexity of the CDF}\label{app:convexity of CDF}

For $\Hat{b}_{t,i}\geq 0$ we investigate the convexity conditions of the Weibull CDF \eqref{eq:WeibullCDF} by the second derivative:
\begin{subequations}
    \begin{align}
        F'(\Hat{b}_{t,i})=f(\Hat{b}_{t,i})&=\kappa\gamma (\Hat{b}_{t,i})^{\gamma-1}\exp(-\kappa (\Hat{b}_{t,i})^{\gamma}), \label{eq: CDF first derivative} \\
        F''(\Hat{b}_{t,i})=f'(\Hat{b}_{t,i})&=\kappa\gamma(\gamma - 1) (\Hat{b}_{t,i})^{\gamma-2}\exp(-\kappa (\Hat{b}_{t,i})^{\gamma})\nonumber \\ 
        &\qquad- \kappa^2\gamma^2 (\Hat{b}_{t,i})^{2(\gamma-1)}\exp(-\kappa (\Hat{b}_{t,i})^{\gamma}). \label{eq:Weibull second derivative} \\
\intertext{\indent $F(\Hat{b}_{t,i})$ is strictly convex for all $\Hat{b}_{t,i}\geq 0$ if $F''(\Hat{b}_{t,i}) > 0$ or strictly concave if $F''(\Hat{b}_{t,i}) < 0$. Canceling common terms in \eqref{eq:Weibull second derivative}, namely $\exp(-\kappa (\Hat{b}_{t,i})^{\gamma})$, $\kappa\gamma$, and $(\Hat{b}_{t,i})^{\gamma-2}$, which are always non-negative, we find the convexity of $F(\Hat{b}_{t,i})$ by the sign of the remaining terms:}
        &\text{sign}(\gamma - 1 - \kappa\gamma (\Hat{b}_{t,i})^{\gamma}).
\intertext{\indent $F(\Hat{b}_{t,i})$ is either convex or concave depending on the sign of $F''(\Hat{b}_{t,i})$ which solely depends on $\gamma$. When $\gamma\leq1$, $F(\Hat{b}_{t,i})$ is globally concave. When $\gamma > 1$, $F(\Hat{b}_{t,i})$ is convex, then concave, with the inflection point:}
    \Hat{b}_{t,i}^* &= \left(\frac{\gamma-1}{\kappa\gamma} \right)^{1/\gamma},
    \end{align}
\end{subequations}
i.e., $F(\Hat{b}_{t,i})$ is convex on $(0, \Hat{b}_{t,i}^*)$ and concave on $(\Hat{b}_{t,i}^*,\infty)$.

\subsection{Convexity of the inverse CDF}\label{app:convexity of inverse CDF}
Let $F:(a,b)\xrightarrow[]{\text{onto}} (c,d)\subset\mathbb{R}$ be a convex function and let $F^{-1}:(c,d)\rightarrow\mathbb{R}$ be its inverse. Then, if $F$ is increasing then $F^{-1}$ is increasing and concave \cite{Mrevi2008}. This intuition also applies in the other direction, from concave to convex. Due to the nature of CDFs, we know that $F$ is an increasing function, and since the function is convex on $(0,\Hat{b}_{t,i}^*)$ and concave on $(\Hat{b}_{t,i}^*,\infty)$, we know that its inverse is concave on $(0,\Hat{b}_{t,i}^*)$ and convex on $(\Hat{b}_{t,i}^*,\infty)$.

\subsection{Uniqueness of the lower-level best response}\label{app:uniqueness stackelberg}
Given an isolated lower-level problem, we analyze the uniqueness of its best response given the TSO's reliability decision $1-\varepsilon_t^*$ in the following sections.
\subsubsection{Reserve providers}
The optimal bid of a stochastic reserve provider is unique and given by $\hat{b}_{t,i}^* = F^{-1}\left(\varepsilon_t^*/0.2\right)$, due to their maximization objective.

\subsubsection{Market-clearing}
The market-clearing problem minimizes cost subject to a demand balance constraint and capacity bounds, yielding a classical merit-order dispatch. From the stationarity condition of the Lagrangian, the optimal price must satisfy:
\begin{subequations}
    \begin{align}
        \lambda_t^* &= c_{t,i}(1-\varepsilon_t^*) + \bar{\nu}_{t,i} - \underline{\nu}_{t,i}, \quad \forall i, \label{eq:stationarity lambda}
\intertext{where $\lambda_t^*$ is the optimal value of $\lambda_t$. Complementary slackness then partitions units into three groups: Infra-marginal units are dispatched at their upper bound $b^*_{t,i} = \hat{b}^*_{t,i}$, implying $\underline{\nu}_{i,t}=0$ and $\bar{\nu}_{t,i}\geq 0$, from which we find:}
    \lambda_t^* &\geq c_{t,i}(1-\varepsilon_t^*).
\intertext{Extra-marginal units are shut down at $b^*_{t,i} = 0$, implying $\underline{\nu}_{i,t}\geq0$ and $\bar{\nu}_{t,i}= 0$, from which we find:}
    \lambda_t^* &\leq c_{t,i}(1-\varepsilon_t^*).
\intertext{The marginal unit $i^*$ clears the residual demand $b^*_{t,i^*} = d_t - \sum_{i < i^*}\hat{b}^*_{t,i}$, i.e. $0 < b^*_{t,i} < \hat{b}^*_{t,i}$, which implies $\underline{\nu}_{t,i}=\bar{\nu}_{t,i}=0$, from which we find:}
        \lambda^*_t &= c_{t,i^*}(1-\varepsilon^*_t).
\intertext{\indent The primal solution $b^*_{t,i}$ is unique but fails when two units share an identical cost at the margin, i.e. when $c_{t,i}(1-\varepsilon^*_t) = c_{t,j}(1-\varepsilon^*_t)$, $i\neq j$, producing a continuum of feasible dispatches along $b^*_{t,i} + b^*_{t,j}$, which we avoid in our case study by design. To see this via the KKTs, note that stationarity in \eqref{eq:stationarity lambda} holds with 
$\bar{\nu}_{t,k}=\underline{\nu}_{t,k}=0, k\in\{i,j\}, i\neq j$ for any convex combination $(b^*_{t,i},b^*_{t,j})$. The demand balance constraint requires only that their sum is fixed, $b^*_{t,i}+b^*_{t,j}=\mathrm{const.}$; complementary slackness is satisfied trivially, so every point on the segment is a KKT point and hence a global optimum. Similarly, the dual variable $\lambda^*_t$ is unique only when the marginal unit is strictly interior, $b^*_{t,i^*} \in (0, \hat{b}^*_{t,i^*})$; if instead demand is met exactly at a capacity bound, stationarity permits}
    \lambda^*_t &\in \left[c_{t,i^*}(1-\varepsilon^*_t),\ c_{t,i^*+1}(1-\varepsilon^*_t)\right],
    \end{align}
\end{subequations}
leaving the price indeterminate within an interval. Thus, both primal dispatch and dual prices are unique, with degeneracy arising from reserve provider costs or boundary solutions, structurally possible outcomes in practice.

\subsection{McCormick envelopes}\label{app:mccormick}
McCormick envelopes is a type of convex relaxation that can be applied to optimization problems that contain bilinear terms of continuous variables. The relevant bilinear terms are $\varepsilon_t b_{t,i}$ in \eqref{eq:TSO objective}, $\bar{\mu}_{t,i}z_{t,i}$ in \eqref{eq:strong duality1}, and $\bar{\nu}_{t,i}\hat{b}_{t,i}$, $\lambda_t d_t$, and $\varepsilon_t b_{t,i}$ in \eqref{eq:strong duality2}. 

Given a bilinear expression $u_{t,i}=(1-\varepsilon_t) b_{t,i}$ where each variable is continuous and has defined lower and upper bounds from \eqref{eq:TSO epsbounds}, \eqref{subeq:supply bounds market clearing} and \eqref{subeq:market LB}, the new objective function with the relaxations are defined as:
\begin{subequations}\label{eq:McCormick optimization}
    \begin{align}
    \min_{\boldsymbol{\xi},\boldsymbol{u}}\quad &\sum_{t,i}\alpha_{t,i}b_{t,i} + \beta_{t,i}u_{t,i} \nonumber \\&\qquad+ 
        \sum_{t,\omega}\pi_{t,\omega}\left( \rho^{\rm{viol}}
        s_{t,\omega}^{\rm viol} + \rho^{\rm{sys}} 
        s_{t,\omega}^{\rm sys}\right) \\
    \intertext{with added constraints:}
        &u_{t,i} \geq 0.8 b_{t,i}, &&\hspace{-1cm} \forall t,i, \\
        &u_{t,i} \geq b_{t,i} -\varepsilon\Hat{b}_{t,i}, &&\hspace{-1cm} \forall t,i,\\
        &u_{t,i} \leq b_{t,i}, &&\hspace{-1cm} \forall t,i,\\
        &u_{t,i} \leq 0.2\Hat{b}_{t,i}-\varepsilon\Hat{b}_{t,i} + 0.8 b_{t,i}, &&\hspace{-1cm} \forall t,i. 
    \end{align}
\end{subequations}
This method approximates the bilinear terms in \eqref{eq:FULL OPT} but simplifies the resulting optimization problem for computational efficiency. The method is applied to all bilinear terms in \eqref{eq:FULL OPT}. For dual variables, bounds are found analytically in preparation for the relaxation (see Appendix \ref{app:dual upper bounds}).

When solving for the optimal static reliability threshold, the McCormick relaxation yields a lower bound on the objective value, which compared to evaluating the original bilinear objective \eqref{eq:TSO objective} at the relaxed solution gives a relaxation gap of up to 4\%, depending on the penalty.

\begin{figure}[t]
    \centering
    \includegraphics[scale=1]{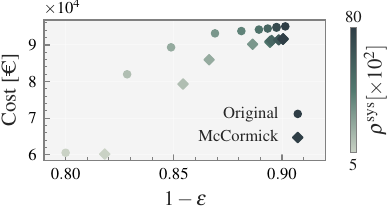}
    \caption{\small{Cost versus reliability for \eqref{eq:FULL OPT} (circles) and \eqref{eq:McCormick optimization} (diamonds). The McCormick relaxation consistently yields lower costs at comparable reliability levels, reflecting the optimistic bias introduced by replacing bilinear terms with their convex envelope approximations.}}
    \label{fig:mccormick gap}
    \vspace{-4mm}
\end{figure}

Figure \ref{fig:mccormick gap} illustrates the difference in the optimal solution using McCormick envelopes and the Pareto-optimal solution obtained from Fig. \ref{fig:pareto different penalty}. Across the explored range of reliability values, the McCormick relaxation consistently achieves lower costs with comparable, but slightly higher, reliability levels. This gap explicitly reflects the looseness introduced by the convex relaxation. The results suggest that while the McCormick relaxation enables efficient optimization, it introduces a systematic bias that should be accounted for when interpreting solution quality.

\subsection{Bounds on dual variables} \label{app:dual upper bounds}
McCormick envelopes require lower and upper bounds on the variables involved. 
We analytically derive bounds on dual variables that are not bounded by definition.

\subsubsection*{Reserve providers}
The upper bounds on the dual variables $\bar{\mu}_{t,i}$ and $\underline{\mu}_{t,i}$ 
are derived by exploiting the KKT stationarity condition together with the 
non-negativity requirements imposed by complementary slackness. Stationarity 
with respect to $\hat{b}_{t,i}$ yields:
\begin{subequations}
\begin{align}
    &\bar{\mu}_{t,i} - \underline{\mu}_{t,i} = 1,
\intertext{and since both multipliers must be non-negative, each is immediately 
bounded by:}
    &0 \leq \bar{\mu}_{t,i}, \underline{\mu}_{t,i} \leq 1.
\end{align}
\end{subequations}

\subsubsection*{Market-clearing}
From standard economic theory, $\lambda_t$ takes the value of the marginal 
reserve provider's cost, i.e., $\lambda_t$ is bounded below by $0$ 
and above by the cost of the most expensive reserve provider at time $t$:
\begin{subequations}
\begin{align}
&0 \leq \lambda_t \leq \max_{i}\{c_{t,i}(1-\varepsilon_t)\}, \label{eq:lambda bounds}
\intertext{which in our case is the dispatchable reserve provider with a cost independent of the reliability level. The upper bounds on $\bar{\nu}_{t,i}$ and $\underline{\nu}_{t,i}$ 
are derived analogously. Stationarity with respect to $b_{t,i}$ gives:}
    &\bar{\nu}_{t,i} - \underline{\nu}_{t,i} = \lambda_t - c_{t,i}(1-\varepsilon_t), \quad \forall i.
\intertext{Rearranging and invoking non-negativity of both multipliers yields:}
    &\bar{\nu}_{t,i} = \lambda_t - c_{t,i}(1-\varepsilon_t) + \underline{\nu}_{t,i} \geq 0, \\
    &\underline{\nu}_{t,i} = c_{t,i}(1-\varepsilon_t) - \lambda_t + \bar{\nu}_{t,i} \geq 0,
\intertext{which, combined with complementary slackness enforcing that at most 
one multiplier per constraint pair is nonzero at optimality, gives the tight bounds:}
    &0 \leq \bar{\nu}_{t,i} \leq \lambda_t - c_{t,i}(1-\varepsilon_t), \label{eq:nu upper bound} \\
    &0 \leq \underline{\nu}_{t,i} \leq c_{t,i}(1-\varepsilon_t) - \lambda_t. \label{eq:nu lower bound}
\intertext{Substituting the upper bound on $\lambda_t$ from \eqref{eq:lambda bounds} 
into \eqref{eq:nu upper bound} and \eqref{eq:nu lower bound} yields the looser 
but more tractable bounds at time $t$:}
    &0 \leq \bar{\nu}_{t,i} \leq \max_i\{c_{t,i}(1-\varepsilon_t)\}, \\
    &0 \leq \underline{\nu}_{t,i} \leq \max_i\{c_{t,i}(1-\varepsilon_t)\}.
\end{align}
\end{subequations}

\subsection{Linearizations of F}\label{app:linearizations of f}
As $F^{-1}(\varepsilon_t/0.2)$ is nonlinear, we linearize it for numerical simplicity. In the feasible and relevant domain of $1-\varepsilon_t\in[0.8, 1]$, we make $n$ evenly spaced cuts and linearize between them, which Gurobi handles on its own. Hence, there is no need to implement SOS2 \cite{Gabriel} or other methods. The definition of $F^{-1}(\varepsilon_t/0.2)$ is such that it may take negative values which is a direct violation of \eqref{eq:unit LL}. The linearization enables us to remove the equality condition in \eqref{eq:unit inverse CDF} which is causing the violation.

\section*{Acknowledgment}
The authors thank Spirii for sharing the EV charging data  used in this study, and Matti Juhani Koivisto (DTU) for providing access to and guidance on the CorRES wind  simulation tool. Thomas Dalgas Fechtenburg (Energinet) is thanked for valuable discussions on the P90 requirement and the Danish FCR-D market structure; any modeling choices inspired by those discussions reflect solely the authors' own interpretation and do not necessarily represent  the views, practices, or policies of Energinet. This manuscript also includes minor corrections assisted by ChatGPT (OpenAI, GPT-5.5 Instant, May 2026 version) and Claude AI (Anthropic, Sonnet 4.6, May 2026 version), used to improve efficiency of Python scripts and to correct  grammatical errors. All intellectual content, interpretation, and  conclusions are the authors' own.

\bibliography{main.bib}% common b

\end{document}